\documentclass[10pt,letterpaper]{article}

\usepackage{cogsci}
\usepackage{hyperref}

\cogscifinalcopy 

\usepackage{pslatex}
\usepackage{apacite}
\usepackage{float} 

\usepackage{graphicx}
\usepackage{color}

\title{How do people incorporate advice from\\ artificial agents when making physical judgments?}

 \author{
    {\large \bf Erik Brockbank$^{\star}$}$^{\dagger1}$, {\large \bf Haoliang Wang$^{\star}$}$^{1}$, {\large \bf Justin Yang}$^{1}$,
    {\large \bf Suvir Mirchandani}$^2$, \\
    {\large \bf Erdem B{\i}y{\i}k}$^{2}$, {\large \bf Dorsa Sadigh}$^{2}$, \and {\large \bf Judith E. Fan}$^1$ \\ 
    $^1$ University of California San Diego,
    $^2$ Stanford University \\
    $^{\dagger}$\texttt{ebrockbank@ucsd.edu}}

\begin{document}

\maketitle

\begin{abstract}
How do people build up trust with artificial agents? Here, we study a key component of interpersonal trust: people's ability to evaluate the \textit{competence} of another agent across repeated interactions. Prior work has largely focused on appraisal of simple, static skills; in contrast, we probe competence evaluations in a rich setting with agents that learn over time. Participants played a video game involving physical reasoning paired with one of four artificial agents that suggested moves each round. We measure participants' decisions to accept or revise their partner's suggestions to understand how people evaluated their partner's ability. Overall, participants collaborated successfully with their agent partners; however, when revising their partner's suggestions, people made sophisticated inferences about the competence of their partner from prior behavior. Results provide a quantitative measure of how people integrate a partner's competence into their own decisions and may help facilitate better coordination between humans and artificial agents.

\end{abstract}

\section{Introduction}

\textit{How do people build up trust across repeated interactions?} This question has motivated research from diverse areas of cognitive science spanning social psychology \cite{simpson2007psychological, deutsch1973resolution} as well as game theory and economics \cite{camerer1988experimental, berg1995trust}. As artificial intelligence agents become increasingly ubiquitous in our everyday lives, the question of how to build up trust with them has also gained prominence in human-computer interaction (HCI) and robotics \cite{soh2020multi, chen2020trust}. 

For instance, in autonomous driving settings people routinely make decisions about how much to trust an artificial driving agent. And in many industrial domains, people work closely with automated agents, sometimes for high stakes tasks. The emergence of trust in our interactions with artificial agents involves a range of complex social inferences, such as recognizing that they share our goals or utilities to begin with \cite{serrino2019finding}. However, one of the central features of human collaboration with artificial agents is that we trust them to be \textit{competent} across a range of task settings. Indeed, greater levels of trust may simply correspond to a belief that the agents are competent in a wider range of settings; for example, trust in an autonomous vehicle may in large part reflect a belief that it can handle a suitably broad array of driving challenges.

How then do people assess another agent's competence over repeated interactions? Prior work in developmental psychology suggests that inferences about another person's competence emerge early in development and draw on a rich set of abstractions about task difficulty and human behavior \cite{gweon2021inferential, leonard2019better}. As adults, this ability continues to develop, allowing us to make complex inferences about other people which draw on rich internal models \cite{velez2019integrating, velez2021learning}, meta-cognitive skills \cite{pescetelli2021role}, and expectations \cite{leong2018unrealistic, chang2010seeing}. Recent work in robotics and HCI suggests that when determining a robot or artificial agent's competence, people may rely on similar cognitive processes, leveraging abstractions about both the agent---e.g., their risk aversion \cite{xie2019robot}---and the environment, such as how much the agent's ability will generalize across tasks \cite{soh2020multi}. 

Despite this convergence of findings across psychology and artificial intelligence, there remain significant challenges in characterizing how people assess the competence of another agent. For one, real-world judgments of competence are often nebulous. How good is somebody at baking or predicting the stock market or writing academic papers? Second, in many complex settings, people's judgments of another agent's competence rely in large part on that agent's ability to \textit{learn} in the task environment. 

The current study builds on prior work by addressing both of these aspects of people's competence evaluations. First, unlike prior work on advice taking that has focused on judgments about abstract variables, e.g., change in stock market prices \cite{leong2018unrealistic} or the outcome of a card flip \cite{velez2019integrating}, here we explore how people incorporate input from an artificial agent when predicting concrete physical events. Second, rather than isolating competence judgments about static agents \cite{chen2020trust}, the current experiment probes people's ability to detect another agent's \textit{learning} over time. A better understanding of how an agent's learning impacts competence judgments in a rich physical domain may lead to more general insights into how people reason about the abilities of others, and how this reasoning impacts their subsequent decisions to trust them in a range of everyday settings.

\begin{figure*}[hbtp]
\vspace{-8mm}
\includegraphics[width=\textwidth]{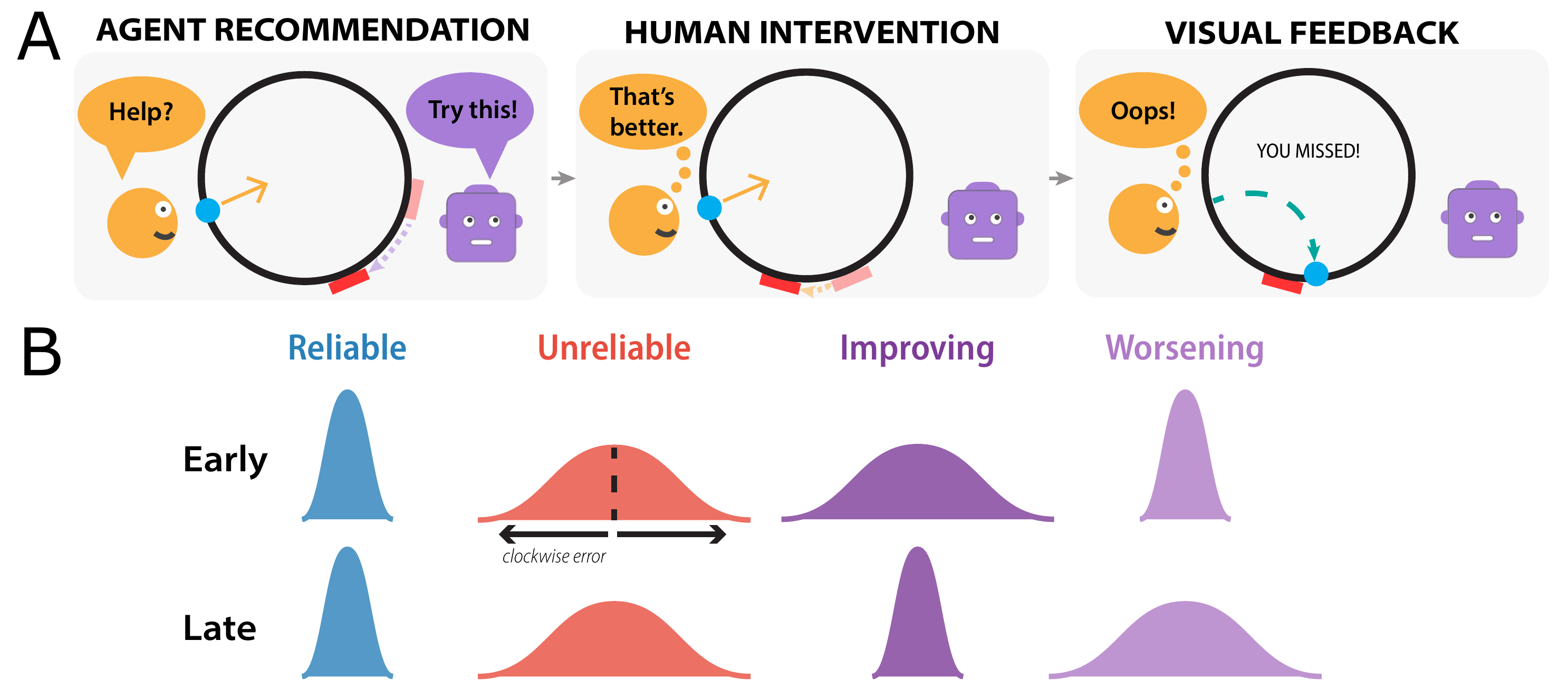} 
\vspace{-4mm}
\caption{Task and Experimental Design. (A) Participants worked with an artificial agent partner to catch a ball launched from the edge of a circle. Their partner began by suggesting a paddle location which participants could either accept or modify. (B) The agents chose suggested paddle locations from a distribution around the ball's true landing position. The variance of this distribution determined how reliable the agent's suggestions were. Participants were assigned to one of four conditions that varied the reliability of the agent's paddle suggestions over the course of the experiment.} 
\label{fig:stimuli}
\end{figure*}

\newpage

Here, we investigate people's ability to collaborate with a dynamic artificial agent in a challenging physics-based video game. Participants were tasked with catching a ball launched from different locations around a circle by placing a paddle where the ball would land. Each round, they were given a suggestion from their agent partner about where to place the paddle to catch the ball. Building on prior work, we use people's decisions about whether to accept or modify their partner's suggestions to probe judgments of the partner's competence \cite{xie2019robot, chen2020trust}. Critically, people's agent partners varied in their true competence and improvement over time. We first ask how much people's behavior in the game draws on their own physical judgments versus the suggestions of their partner and how this varies based on their partner's competence. Next, we ask whether people's intervention decisions reflect ongoing assessments of their partner's ability rather than trial-specific context. 
Our experiments revealed several key findings: First, rather than relying exclusively on their own physical judgments or the advice of their partner, people integrated both sources of information in their interventions. Moreover, the degree to which they incorporated their partner's input was predicted by how reliable the agent had been in the past, not just the quality of its current advice. Taken together, our results suggest that people's physical judgments in collaborative settings involve rich, abstract inferences about others based on their past behavior.

\section{Experiment}

\subsection{Participants}

256 adults recruited from Prolific completed the task online. Data from 12 participants were excluded from subsequent analyses due to technical issues encountered during the experiment, resulting in 244 participants with complete data (average age: 33.8 years, $SD$ = 11.3; 127 male, 103 female, 13 non-binary; educational background distributed across high school, 4-year college, and graduate degrees). The experiment lasted approximately 25 minutes and participants were paid \$14/hr based on this expected completion time. 
All participants provided informed consent in accordance with the UC San Diego IRB.

\subsection{Human-agent collaboration task}

In the experiment, participants tried to catch a virtual ball launched from a point on a circle using a rectangular paddle positioned along the outside of the circle (see Figure \ref{fig:stimuli}).\footnotemark{} 
Participants worked together with an artificial agent ``partner'' who was trying to help them on the task. On each round, the partner suggested a paddle location based on the ball's launch position; participants could either accept this suggestion or adjust the paddle themselves before launching the ball.

\footnotetext{All code used to run the experiment, as well as code used in analyses below, can be found at: 

\url{https://github.com/cogtoolslab/
trust\_agents\_cogsci2022\_public}.}

Each trial began with participants' agent partner suggesting a paddle location that would catch the ball; the paddle was shown moving around the circle and a small animation on the right showed the agent ``thinking.'' Once the agent had moved the paddle to its suggested location, participants were given the opportunity to either adjust the paddle with the arrow keys or keep their partner's suggestion. If participants adjusted the paddle, the agent's original recommendation remained visible and marked in gray. When participants settled on a paddle location, they launched the ball with the spacebar. The ball's path was animated and participants were shown a message indicating whether they had successfully caught it before proceeding to the next trial. 

Every session consisted of 96 trials divided into eight blocks of 12. These ``blocks'' were not visible to participants; in each block, the ball appeared at locations sampled in a random order from each of 12 bins of equal width along the circle's circumference.

\subsection{Manipulating agent ability}

Participants were assigned to one of four conditions that manipulated the quality of their partner's suggested paddle locations: an \textit{unreliable} partner, a \textit{reliable} partner, an \textit{improving} partner, and a \textit{worsening} partner. The agent's suggested paddle location on each trial was an angle $x$ sampled from a \textit{von Mises} distribution (a circular approximation to a normal distribution) with mean $\mu$ equal to the ball's final landing angle $\rho$, and variance $\sigma^2$ determined by the agent's competence level. The \textit{reliable} agent had a low $\sigma^2 \approx 10$ degrees; the sampled paddle location was almost always close to the ball's true landing location. By contrast, the \textit{unreliable} agent sampled its paddle locations from a high-variance distribution with $\sigma^2 \approx 48$ degrees. The high and low-competence $\sigma^2$ values were chosen to give the agents expected success rates of around $80\%$ and $20\%$, respectively. Meanwhile, the \textit{improving} agent began with a $\sigma^2$ value equal to the \textit{unreliable} agent's but every 12 trials the variance decreased by a fixed amount so that during the final 12 trials, it had a $\sigma^2$ equal to the \textit{reliable} agent's. The \textit{worsening} agent was symmetrical but in the opposite direction.

\subsection{Measuring human appraisals of agent ability}

A core goal of our study was to investigate the impact of manipulating an agent's behavior on participants' impressions of its competence, thereby impacting how they approached collaborating with it.
We measured participants' appraisals of their partner's task ability as the degree to which they intervened before committing to a final paddle location on each trial. Intuitively, participants who judged their partner to be more competent would be less likely to revise their partner's suggestion, or do so to a lesser extent. On each trial, we measured whether participants intervened to adjust the paddle's position away from their partner's initial suggestion and the magnitude of this intervention.

If participants were maintaining an ongoing estimate of their partner's task competence, their intervention behavior might be guided by this estimate above and beyond the trial-specific accuracy of their partner's suggestions. For example, participants might place more confidence in the suggestions of the \textit{reliable} agent relative to the \textit{unreliable} agent, even when equating the magnitude of the error in the agent's current recommendation. To isolate the impact of learned expectations about each agent's ability on participants' interventions, we included a \textit{critical trial} in each 12-trial block (unbeknownst to participants): Rather than sampling locations as described above, the suggested paddle location on critical trials was set to a fixed distance from the ball's landing location that was close to the true landing location (approximately 16 degrees) yet would result in missing the ball unless the participant intervened. Including these critical trials enabled direct comparisons between conditions while controlling for the magnitude of the error in the agent's initial suggestion.

\subsection{Post-study questionnaire}

After completing all 96 trials, participants were given a post-study questionnaire to collect basic demographic information and two additional variables we did not analyze here: prior physics courses taken and prior experience with video games. Next, they were asked how often they thought they had intervened on the previous trials and how often they would expect to intervene if they were to play another 96 rounds with this same partner (both 1-100\% scales). Finally, they were asked to indicate how much they trusted the agent to catch the ball (five-point rating scale) and to describe how they decided whether to intervene in the task.

\section{Results}
We began by examining the performance of human-agent teams on the task overall. They caught the ball on 73.8\% (SD = 14.7\%) of trials across all conditions, improving from 55.9\% in the first trial block to 82.8\% in the final block. The root mean squared error (RMSE) of the final paddle locations was 14.85 degrees (SD = 7.56 degrees). Together, these findings suggest that while the task was challenging, participants were nevertheless able to achieve reasonably high performance with their agent partners. However, our primary interest is in how their behavior differed across conditions as a result of differences in their \textit{partner's} ability.

\subsection{People combine information sources to make intervention decisions}

To understand how participants coordinated with their partner, we compare three possible accounts: First, it may be that people trusted their agent partner completely, regardless of its competence. On this view, participants' own physical intuitions would have played no role in their decisions. A second account takes the opposite perspective; people may have ignored their partner's suggestions, simply choosing the best paddle position each round (i.e., if the agent's suggestion was accurate, people would accept it and if not, they would intervene to correct it). Finally, a third possibility is that people's behavior was somewhere in the middle of these two. Rather than consistently following their partner's suggestion or unilaterally seeking the optimal paddle position each round, people may have relied on a combination of their own physical intuitions and their partner's recommendation to decide where to place the paddle. We consider each of these options below; our results suggest that participants integrated intuitive physical judgments with their partner's guidance and that \textit{how much} they incorporated their partner's suggestions was calibrated to the their partner's task performance.

\subsubsection{Participants intervened to improve accuracy} 

\begin{figure}[t]
\begin{center}
\includegraphics[width=0.9\linewidth]{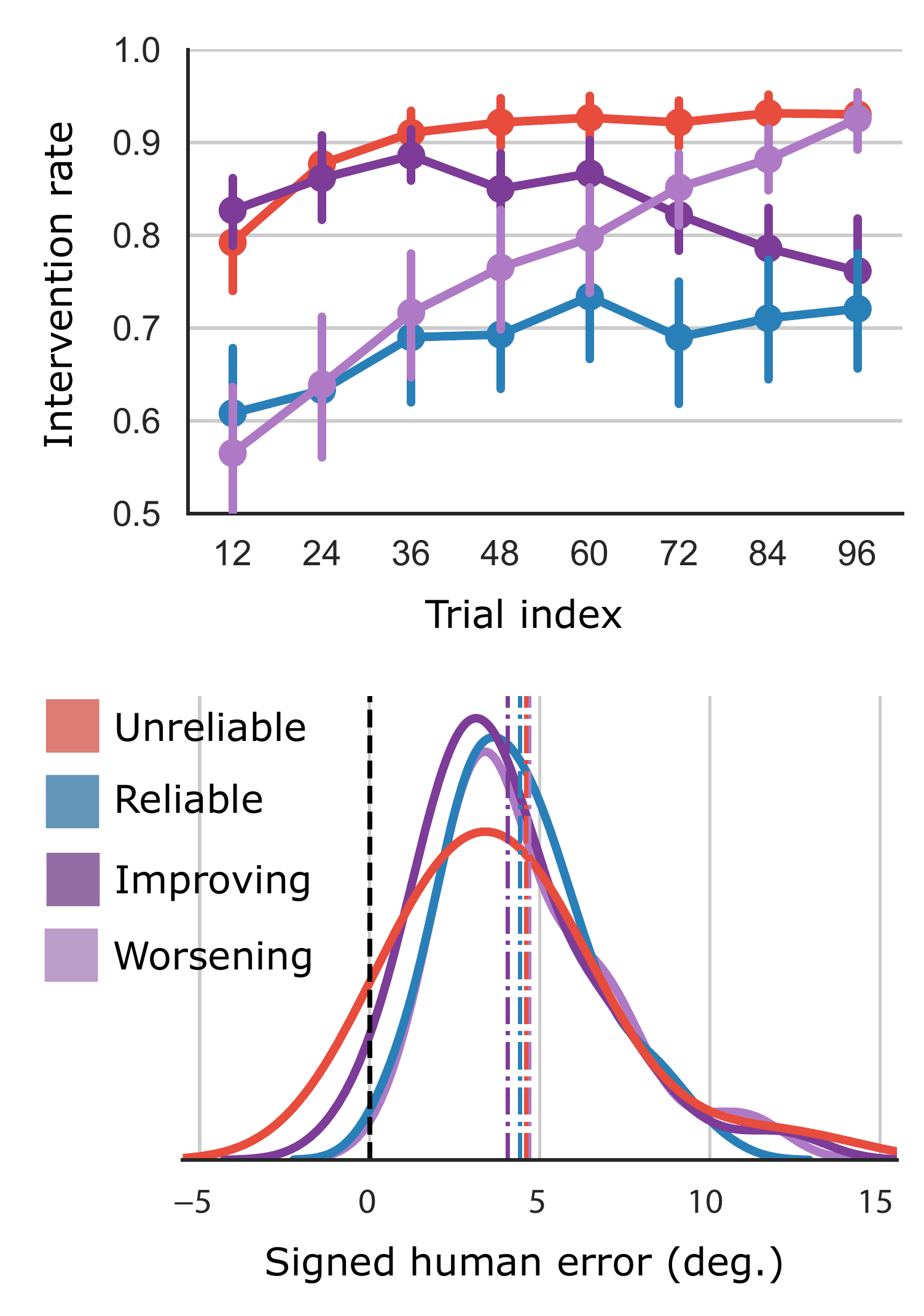}
\end{center}
\caption{
Top: Mean paddle intervention rates by block. 
Bottom: Error distribution in each condition, with positive values indicating responses whose error was in the same direction as the agent's suggestion and negative values indicating the opposite. The black dashed line indicates the true landing location of the ball. Colored vertical lines indicate means in each condition. Curves are based on kernel density estimation.}
\label{fig:interventions}
\end{figure}

We start by considering the first hypothesis above, that people merely acted in accordance with their partner's suggestions. If this were true, we would expect intervention rates to be low and performance in each condition to closely match the ability of the agents in that condition. Figure \ref{fig:interventions} (top) shows average intervention rates (the percent of trials in which each subject modified the agent's original suggestion) in each trial block. Notably, intervention rates were high in all conditions, even with the \textit{reliable} agent, whose suggestions would catch the ball on approximately 80\% of trials. 

Figure \ref{fig:interventions} shows an overall increase in intervention rates even in the \textit{reliable} and \textit{unreliable} conditions where agent performance did not change. This seems most likely to be a result of participants' general task improvement noted above. A generalized linear mixed effects model fit to participants' intervention decisions (binary) with a random intercept for each participant showed a significant main effect of trial block ($\chi^2(1) = 226.2$, $p < 0.001$) and a significant interaction between trial block and condition ($\chi^2(3) = 404.8$, $p < 0.001$). Thus, far from merely trusting their partner's suggestions, participants took an active role in intervening and calibrated their interventions to their partner's underlying ability.

\subsubsection{Intervention decisions incorporated agent suggestions}

In light of the high intervention rates across conditions, one account of people's behavior is that they simply relied on their own intuitive physics to respond. On this view, the quality of their partner's recommendations would have been irrelevant. 

To test this possibility, we examine the distribution of \textit{errors} on each trial relative to both the ball's final landing location and the agent's suggestion. Intuitively, if people completely disregarded their partner's suggestion, their errors would be centered on the ball's true landing location. Alternatively, if people took their partner's suggestion into account, we might expect final paddle placements to be systematically biased towards or away from the partner's initial suggestion. 

Figure \ref{fig:interventions} (bottom) shows the distributions of participants' average error in each condition. Critically, these distributions are signed relative to the ball's landing location and the agent's paddle suggestion; error greater than 0 represents participants placing the paddle away from the ideal catching location \textit{in the direction of the agent's suggestion}. Meanwhile, error less than 0 represents participants placing the paddle away from the ideal location \textit{in the opposite direction of the agent's suggestion}. The dashed lines in Figure \ref{fig:interventions} (bottom) show the average signed error in each condition. 

Participants' signed error was significantly greater than $0$ in all four conditions, reflecting a stable bias toward their partner's recommended paddle locations (\textit{reliable}: $t(56) = 15.70$; \textit{improving}: $t(64) = 12.34$; \textit{worsening}: $t(54) = 14.20$; \textit{unreliable}: $t(66) = 8.13$, all $p$s $< 0.001$). This suggests that people's decisions about where to place the paddle were not merely an effort to find the best location independent of their partner's advice; rather, they showed a systematic anchoring towards the agent's recommendation.

Taken together, the results in Figure \ref{fig:interventions} suggest that people's decisions about where to place the paddle integrated multiple sources of information. They did not merely trust their agent partner regardless of its competence, nor did they simply choose the best move each round without consideration for their partner's recommendation. However, the agent's suggestion on a given trial is not the only source of information that might help participants decide where to ultimately place the paddle. Across repeated interactions, agents in each condition offer ongoing evidence of their underlying \textit{competence} through the accuracy of their paddle suggestions. Participants can use this information to calibrate how much their final paddle locations should be influenced by their partner.

\subsection{People relied on past performance to guide interventions}

Since agent partners varied across conditions in how helpful their paddle suggestions were, we hypothesize that participants incorporated this information into their decisions about how closely to follow their partner's suggestions. 

To test this, we begin by looking at the relationship between the agent's paddle suggestion error and participants' paddle intervention magnitude across conditions. If participants were correcting for the agent's errors in a way that did not integrate the agent's underlying ability, this relationship should be similar across conditions (i.e., they should adjust for small errors less and larger errors more in a similar fashion). We fit a linear mixed effects model of participant intervention magnitude (on trials in which they intervened) as a function of agent recommendation error and condition with a random intercept for participants. Here, we find a significant interaction of condition and agent error ($\chi^2(3) = 293.8$, $p < 0.001$), suggesting that the agent's competence played a critical role in people's \textit{intervention} magnitudes across different \textit{suggestion error} magnitudes. However, this result could be driven in part by the fact that the underlying distribution of agent errors differed substantially across conditions (by design). Thus, a more apples-to-apples comparison should examine people's intervention behavior for similar levels of agent error across conditions. For this, we turn to the eight \textit{critical trials} that each participant completed.

\subsubsection{Critical trial interventions reflected differences in agent ability}

\begin{figure}[h]
\begin{center}
\includegraphics[width=0.9\linewidth]{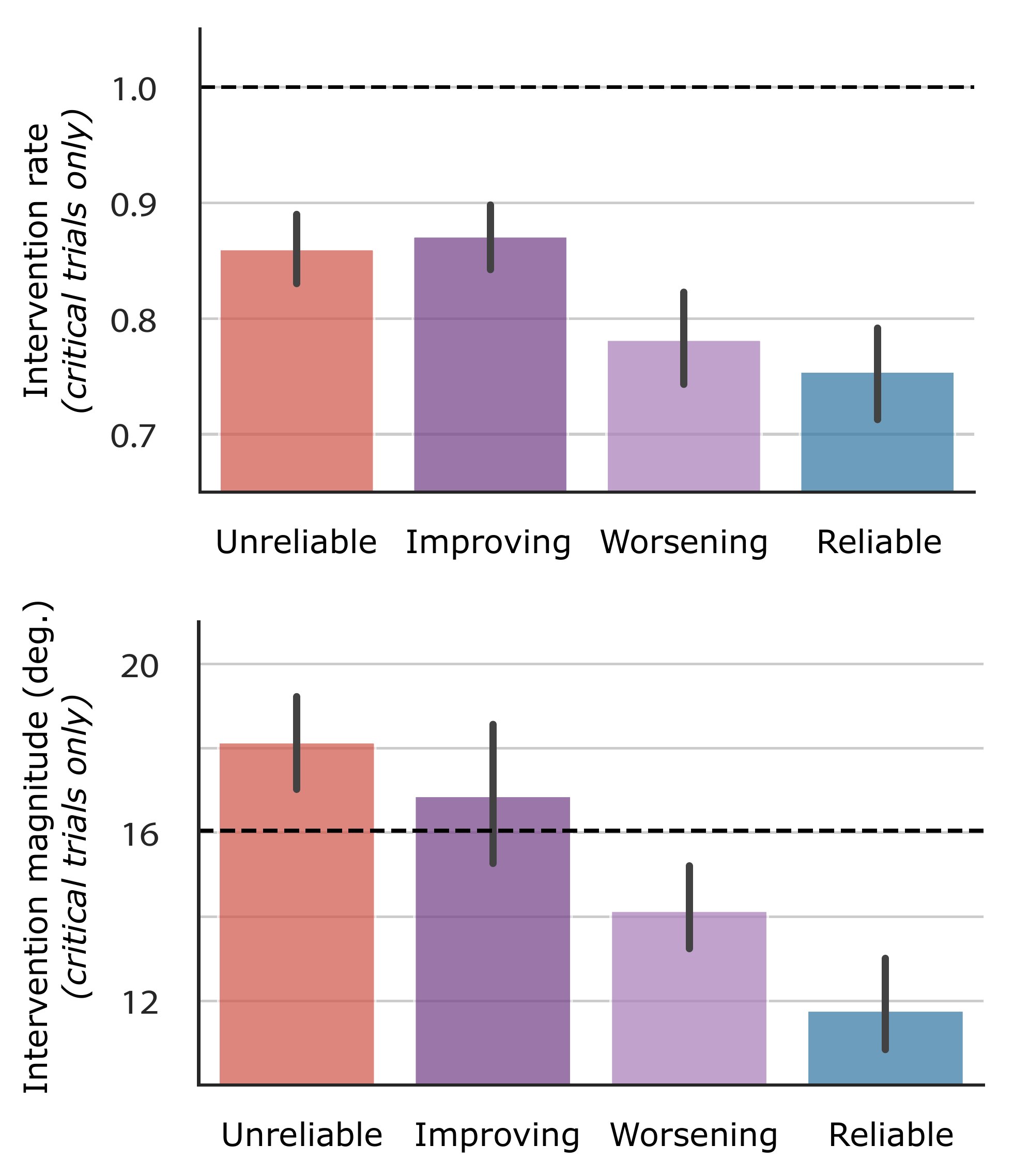}
\end{center}
\caption{Intervention behavior on \textit{critical trials}. 
Top: average proportion of critical trials on which participants chose to intervene. The dashed line indicates optimal behavior (critical trials always required intervention to catch the ball). 
Bottom: average distance participants intervened on critical trials in which they chose to intervene. The dashed line indicates the optimal intervention distance on these trials.} 
\label{fig:critical_trials}
\end{figure}

If people's responses combined their own estimate of the ball's final location and their partner's suggestion---without regard to their partner's overall reliability---we should not see any difference in intervention behavior on the critical trials, since the agent's paddle suggestion error on critical trials was the same across conditions. Figure \ref{fig:critical_trials} (top) shows average intervention rates on critical trials. We fit a generalized linear mixed effects model to participants' interventions (binary) on critical trials; the fixed effect of condition produces a significantly better fit than random intercepts and correlated slopes to account for individual increases in intervention rate over the experiment ($\chi^2(3) = 16.5$, $p < 0.001$). Consistent with the pattern observed in Figure \ref{fig:critical_trials} (top), estimated marginal means were significantly different between \textit{unreliable} and \textit{reliable} conditions ($p = 0.01$), as well as \textit{improving} and \textit{reliable} ($p = 0.003$). Thus, decisions about when to intervene on critical trials were sensitive to differences in the agents' underlying abilities.

Participants' decisions about \textit{how much} to intervene on critical trials (Figure \ref{fig:critical_trials}, bottom) shows a similar pattern. While those paired with an \textit{unreliable} or \textit{improving} partner adjusted the paddle by an amount close to the optimal level, participants whose partner was very accurate (\textit{reliable}) or started out highly accurate (\textit{worsening}) made smaller adjustments on critical trials. In a linear mixed effects model predicting intervention distance---on critical trials where participants intervened---we found a significant effect of condition relative to a baseline model which included only random effects of subject ($\chi^2(3) = 28.3$, $p < 0.001$). Estimated marginal means were significantly different across \textit{unreliable} and \textit{reliable} agents ($p < 0.001$), \textit{unreliable} and \textit{worsening} ($p = 0.005$), and \textit{improving} and \textit{reliable} ($p < 0.001$). Thus, a complete account of reasoning on this task suggests that people maintain an underlying assessment of their partner's competence over time and calibrate their decisions about whether to intervene, and how much, based on this assessment.

Notably, both intervention rates and magnitudes in Figure \ref{fig:critical_trials} show a similar imbalance between intervention behavior with the \textit{improving} and \textit{worsening} agents, despite the fact that these two exhibited symmetrical patterns of learning and deteriorating performance. This raises the intriguing possibility that participants accorded more weight to earlier trials when evaluating the competence of these dynamic partners. However, further work is needed to confirm this.

\begin{figure}[htbp]
\begin{center}
\includegraphics[width=0.9\linewidth]{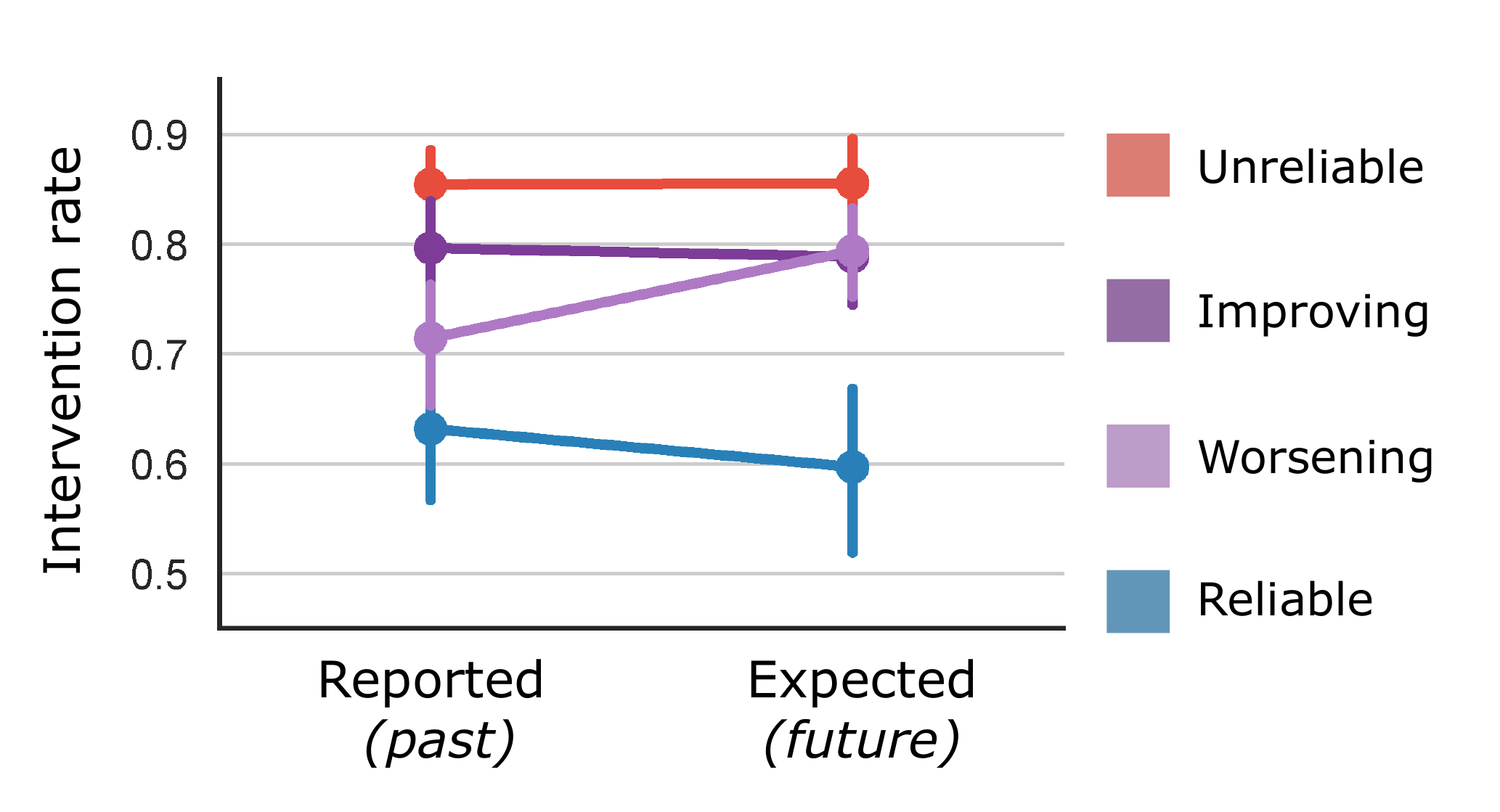}
\end{center}
\caption{Participant estimates of how much they intervened with each agent compared to how much they would expect to intervene in future rounds with the same partner.} 
\label{fig:survey}
\end{figure}

\subsection{Past performance and future expectations}

So far, we have found evidence that participants' decisions about when to intervene with their partner, and how much, rely on an ongoing estimate of their partner's overall competence and that such estimates are sensitive to differences in ability and learning rate over time. \textit{But how predictive were these estimates?} To better understand the granularity of participants' judgments of their partner's competence, we examine responses on the post-experiment survey. 

Figure \ref{fig:survey} shows participants' subjective estimates of how often they had intervened in the experiment trials, paired with their responses indicating how much they would \textit{expect} to intervene if they played 96 more rounds with the same partner. These results provide an estimate of the accuracy of participants' expectations about their partner based on past performance. Responses suggest that people's ability to explicitly forecast their partner's future behavior was limited. In a 4 (between-subjects condition: \textit{reliable}, \textit{unreliable}, \textit{improving}, \textit{worsening}) by 2 (within-subjects rating type: \textit{reported}, \textit{expected}) repeated measures ANOVA of the intervention rates shown in Figure \ref{fig:survey}, differences between conditions were significant ($F(3, 240) = 17.16$, $p < 0.001$) and the interaction between condition and rating type was significant ($F(3, 1) = 6.77$, $p < 0.001$). However, in follow-up paired $t$-tests, participants in the \textit{worsening} condition showed a significant difference between past and expected future intervention rates, implying some degree of forecasting ($t(54) = -4.80$, $p < 0.001$), but participants in the \textit{improving} condition did not ($t(64) = 0.38$, $p = 0.71$). Forecasted intervention rates remained stable for the \textit{unreliable} and \textit{reliable} agents, as we might expect (\textit{unreliable}: $t(66) = -0.10$, $p = 0.92$; \textit{reliable}: $t(56) = 1.58$, $p = 0.12$). 

Critically, forecasted intervention rates reflect participants' underlying trust in their partner. Predicted intervention rates ($0-100\%$) were significantly negatively correlated with responses on a five-point rating scale question asking how much participants trusted their partner to catch the ball on a given trial (\textit{Not at all}, \textit{Slightly}, \textit{Moderately}, \textit{Very}, \textit{Extremely}), $r = -0.51$, $p < 0.001$. This highlights the potential role of expectations about future behavior in our trust in others.

\section{Discussion}

In this study, we address the question of how people evaluate an artificial agent's competence in a collaborative physical prediction task. Specifically, we investigated how differences in an agent's competence impacted people's decisions to either trust their partner's recommendation or intervene to modify it. The current task expands on prior work by probing people's sensitivity to changes in their partner's ability in a setting that draws on rich human physical intuitions. Our results contain several key findings. First, we show that participants combine their own physical judgments and the recommendations of their partner. Second, we show that how people integrate their partner's recommendation does not reflect a simple bias towards their partner's estimate; people calibrate \textit{how much to defer to their partner} based on the prior reliability of their partner's suggestions. We further show that this estimate is not static, but rather sensitive to changes in their partner's competence over time. In sum, our findings make headway towards a better understanding of the behavioral underpinnings of trust in artificial agents across repeated interactions. 

The current results raise a number of questions about trust and evaluations of others' competence that merit further investigation. First, how does the information provided by the agent partner impact task performance relative to learning \textit{without another agent}? A comparison of the current results to people's behavior in the absence of any form of social inference will not only provide a baseline for understanding how people integrate the agent's behavior while learning, but could also allow for future work aimed at optimizing the effectiveness of \textit{pedagogical agents}. 

Second, the current results suggest that people integrate their own judgment with their partner's suggestions in a way that is sensitive to their partner's ability, but leave largely unanswered \textit{how} people evaluate their partner or integrate this evaluation into their own behavior. Future work should compare computational models that embody distinct hypotheses concerning the learning mechanisms and/or the relative contributions of social and task-specific information \cite{parnamets2020integration}. In addition, modifications to the task which allow for more flexible behavior might improve understanding of the tradeoffs people make in their own decisions when collaborating with others, e.g., how much to merely imitate their partner or draw on a richer internal model of their partner's behavior \cite{charpentier2020neuro}. 

Finally, a promising avenue for future work would be to further explore the inferences that people draw when other agents display richly structured patterns of errors in more complex physical task domains. For example, when do people infer that another agent may possess one skill but lack another when both are needed to successfully perform a physical task (e.g., aiming a basketball vs. applying enough force when shooting it)? By providing a more thorough account of how people infer the underlying mechanisms that give rise to the behavior of other agents and use these inferences to guide their own actions, such work can advance our understanding of the basis of trust and lead to algorithms that support improved collaboration between humans and artificial agents.

\newpage

\section{Acknowledgments}
EB$^{1}$, JEF, and DS are supported by an ONR Science of Autonomy award. 
JEF is additionally supported by NSF CAREER \#2047191 and a Stanford Hoffman-Yee grant. 
EB$^{2}$, SM, and DS are supported by an ONR YIP Award and an AFOSR YIP Award.


\bibliographystyle{apacite}
\setlength{\bibleftmargin}{.125in}
\setlength{\bibindent}{-\bibleftmargin}
\typeout{} 
\bibliography{main-revisions}

\end{document}